\documentclass[twocolumn,showpacs,amsmath,amssymb,aps,prl,superscriptaddress,tbtags]{revtex4-1}

\usepackage[]{graphicx}
\usepackage[]{verbatim}
\usepackage[]{color}
\usepackage{overpic}
\usepackage{rotating}
\usepackage{mathtools}
\usepackage{appendix}
\usepackage{ulem}
\usepackage[margin=0.6in]{geometry}
\usepackage{gensymb}
\usepackage{tabularx}

\usepackage[]{hyperref}

 \hypersetup{
 colorlinks = true,
 linkcolor = blue,
 urlcolor = blue,
 citecolor = blue}

\usepackage{array}
\newcolumntype{L}[1]{>{\raggedright\let\newline\\\arraybackslash\hspace{0pt}}m{#1}}
\newcolumntype{C}[1]{>{\centering\let\newline\\\arraybackslash\hspace{0pt}}m{#1}}
\newcolumntype{R}[1]{>{\raggedleft\let\newline\\\arraybackslash\hspace{0pt}}m{#1}}
\newcolumntype{N}{@{}m{0pt}@{}}

\makeatletter
\newsavebox{\@brx}

\newcommand{\llangle}[1][]{\savebox{\@brx}{\(\m@th{#1\langle}\)}%
  \mathopen{\copy\@brx\mkern2mu\kern-0.8\wd\@brx\usebox{\@brx}}}
\newcommand{\rrangle}[1][]{\savebox{\@brx}{\(\m@th{#1\rangle}\)}%
  \mathclose{\copy\@brx\mkern2mu\kern-0.8\wd\@brx\usebox{\@brx}}}

  \newcommand{\lllangle}[1][]{\savebox{\@brx}{\(\m@th{#1\langle}\)}%
  \mathopen{\copy\@brx\copy\@brx\mkern4mu\kern-0.7\wd\@brx\usebox{\@brx}}}
\newcommand{\rrrangle}[1][]{\savebox{\@brx}{\(\m@th{#1\rangle}\)}%
  \mathclose{\copy\@brx\copy\@brx\mkern4mu\kern-0.7\wd\@brx\usebox{\@brx}}}

\graphicspath{{./}{./}}

\begin{document}
\title{Distinct reduction of Knight shift in superconducting state of Sr$_2$RuO$_4$ under uniaxial strain}
\author{Austin W.~Lindquist}
\affiliation{Department of Physics and Center for Quantum Materials, University of Toronto, 60 St.~George St., Toronto, Ontario, M5S 1A7, Canada}
\author{Hae-Young Kee}
\email{hykee@physics.utoronto.ca}
\affiliation{Department of Physics and Center for Quantum Materials, University of Toronto, 60 St.~George St., Toronto, Ontario, M5S 1A7, Canada}
\affiliation{Canadian Institute for Advanced Research, Toronto, Ontario, M5G 1Z8, Canada}
\begin{abstract}
Shortly after the discovery of superconductivity in Sr$_2$RuO$_4$, spin-triplet pairing was proposed and further corroborated by 
a constant Knight shift ($K$) across the transition temperature (T$_c$).
However, a recent experiment
observed a drop in $K$ at T$_c$ which 
becomes larger under uniaxial strain, ruling out several spin-triplet scenarios.
Here we show that even-parity inter-orbital spin-triplet pairing can feature a d-vector that rotates when uniaxial strain is applied, leading to a larger drop in the spin polarization perpendicular to the strain direction, distinct from spin-singlet pairing. 
We propose that anisotropic spin polarization under strain 
will ultimately differentiate triplet vs.~singlet pairing. 
\end{abstract}
\maketitle

{\it Introduction} -- The discovery of superconductivity in Sr$_2$RuO$_4$ \cite{Maeno1994Nature} has had great attention over the past two decades.
It has been considered the best solid-state system which exhibits 
a time-reversal symmetry breaking p-wave spin-triplet pairing analog of the A-phase in $^3$He \cite{Mackenzie2003RMP}. 
The microscopic route to the spin-triplet pairing in $^3$He is ferromagnetic fluctuations \cite{Leggett1975RMP}.
 Since a sister compound, SrRuO$_3$, is a ferromagnetic metal, the $p+ip$ spin triplet pairing proposed by Rice and Sigrist \cite{Rice1995JPCM} was a promising candidate.
Earlier experiments of nuclear magnetic resonance (NMR) and $\mu$SR had corroborated this proposal, because no change in the NMR Knight shift \cite{Ishida1998Nature} and
a broken time reversal symmetry signal across $T_c$ in $\mu$SR \cite{Luke1998Nature} 
are consistent with the order parameter.
However, a scanning magnetic imaging \cite{Bjornsson2005PRB}
measurement
showed a null signal of the associated chiral super-current, which
does not support the chiral p-wave spin-triplet pairing. Since then, the pairing symmetry of Sr$_2$RuO$_4$ has remained a mystery with controversial experimental results \cite{Nelson2004,Jang2011,Kallin2012RPP,Mackenzie2017NPJ}.

\begin{figure}
\includegraphics[width=1.0\columnwidth]{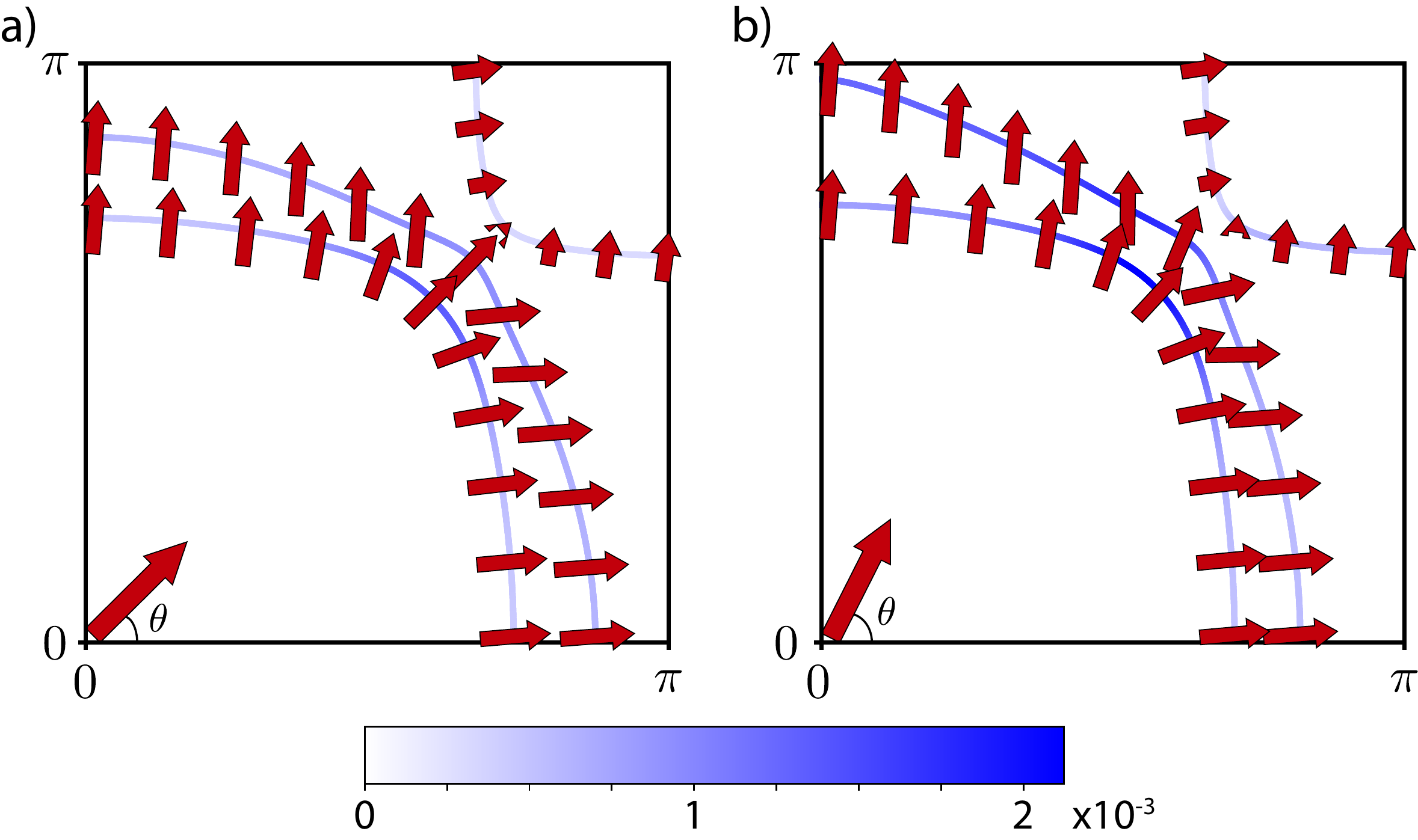}
\caption{The red arrows at representative momenta 
show the d-vector of inter-orbital spin triplet pairing for (a) unstrained and (b) uniaxial-strain along $a$-axis. This transforms to intra-band pseudospin-singlet pairing on the FS and inter-band pairings (see the main text for details).
The d-vector rotation occurs the most in the diagonal direction of the Brillouin zone. 
The length of each arrow represents the in-plane component; the shorter the arrow, the bigger the $c$-axis component.  Note that the arrows with an inverted tail correspond to a vector primarily along the $c$-axis.
The blue colour on the FS denotes the size of gap.
The red arrow at the bottom corner of each panel represents the averaged d-vector direction projected onto the $ab$-plane
denoted by $\theta$; $\theta= 45$ and  63 degrees in (a) and (b) respectively.  }\label{fs}
\end{figure}

Recently, Pustogow et al.~\cite{Pustogow2019Nature} made an important breakthrough in determining the spin component of the order parameter, as they
reported a 20-50\% drop, depending on the field strength, in the spin polarization $M_s$ below $T_c$ in unstrained samples in contrast to the earlier NMR reports \cite{Ishida1998Nature}.
When the sample is strained along the $a$-axis, the spin polarization along the $b$-axis
 drops almost 75\%. This rules out the d-vector along the $c$-axis as in the chiral pairing proposal \cite{Rice1995JPCM}. 
Since the no-change Knight shift across $T_c$ has been the strong piece of evidence of a spin-triplet, 
this observation may rule out several spin-triplet pairings including the d-vector along the $c$-axis, and potentially in the $ab$-plane, depending on the magnitude of the decrease observed, as listed in Ref.~\onlinecite{Pustogow2019Nature}.

Here we show that orbital-singlet spin-triplet (OSST) pairings \cite{Puetter2012EPL} exhibit a significant reduction in the spin polarization under strain, and it becomes anisotropic relative to the strain direction.  
For OSST pairings the d-vector is locked in momentum space
via spin-orbit coupling (SOC) as shown in Fig.~\ref{fs}.
When the uniaxial strain is applied, the strength of the pairing is enhanced due to the van Hove singularity (vHS),
which is true for both spin-singlet and -triplet. However, for an OSST with an in-plane d-vector,  there is an important additional effect of uniaxial strain. It not only enhances the magnitude
of the pairing, but also {\it rotates}  the direction of the d-vector as shown in Fig.~\ref{fs}(b), because the strain changes the composition of orbitals which then affects the d-vector direction.
The d-vector rotation creates an anisotropy between spin polarizations parallel vs.~perpendicular to the strain direction. 
When the strain is applied along the $a$-axis, the d-vector rotates towards the $b$-axis as shown by the red arrows in Fig.~\ref{fs}, leading to
 a larger drop in the spin-polarization along the $b$-axis than that of the unstrained case, as reported in  Ref.~\onlinecite{Pustogow2019Nature}.
With the same strain condition, the $a$-axis polarization drop should be smaller. For a singlet, the two spin polarizations are the same.
Thus we propose a NMR Knight shift experiment with the reasonably large field (but below the 1.5 T upper critical field) along the $a$-axis
under the $a$-axis strain, to be compared with the $b$-axis polarization. This will ultimately differentiate spin-triplet vs.~-singlet pairings.

Below we formulate the proposed idea using a model which consists of a Kanamori interaction and a t$_{2g}$ tight binding model with SOC.
While the atomic SOC leading to an s-wave gap is used for clarity, it can be generalized by including momentum dependent SOC terms leading to any even-parity OSST pairing (such as d-wave or g-wave).

{\it Microscopic Hamiltonian} -- Sr$_2$RuO$_4$ is a multi-orbital system with non-negligible SOC.
The orbital degrees of freedom allow for four distinct pairings which satisfy 
the antisymmetric fermion wave function requirement, i.e., ${\hat \Delta}({\bf k)} = - {\hat \Delta}^T (-{\bf k})$. The four types are:
(i) even-parity intra-orbital (or inter-orbital-triplet) spin-singlet ($\hat{\phi}_a$ or $\hat{\phi}_\nu$), (ii) odd-parity inter-orbital-singlet spin-singlet  
(iii) odd-parity intra-orbital (or inter-orbital-triplet) spin-triplet ($\hat{\textbf{d}}_a$), and (iv) even-parity OSST ($\hat{\textbf{D}}_{\nu}$),
where $\nu$ represents inter-orbital, and $a$, intra-orbital pairings among $t_{2g}$ \cite{Puetter2012EPL}.

A generic Hamiltonian $H=H_{\text{kin}}+H_{\text{SOC}}+H_{\text{int}}$ consisting of a tight binding model, SOC, and Kanamori interaction is considered.  The tight binding and SOC terms are used to reproduce the Fermi surface (FS) reported in Ref.~\onlinecite{Tamai2019PRX}, and are listed in the Supplemental Material (SM).  The underlying FS of three bands, $\alpha$, $\beta$, and $\gamma$ is reported earlier \cite{Mackenzie1996PRL,Bergemann2000PRL,Damascelli2000PRL}, 
and was further refined in Ref.~\onlinecite{Tamai2019PRX} shown as the solid lines in Fig.~\ref{fs}.  The interaction term is given by
\begin{equation}
\begin{aligned}
H_{\text{int}} = \frac{U}{2}\sum_{i,a} c_{i\sigma}^{a\dagger}c_{i\sigma'}^{a\dagger}c_{i\sigma'}^{a}c_{i\sigma}^{a} + \frac{V}{2} \sum_{i,a \ne b} c_{i\sigma}^{a\dagger}c_{i\sigma'}^{b\dagger}c_{i\sigma'}^{b}c_{i\sigma}^{a} \\
+ \frac{J_H}{2}\sum_{i,a \ne b} c_{i\sigma}^{a\dagger}c_{i\sigma'}^{b\dagger}c_{i\sigma'}^{a}c_{i\sigma}^{b}+ \frac{J_H}{2}\sum_{i,a \ne b} c_{i\sigma}^{a\dagger}c_{i\sigma'}^{a\dagger}c_{i\sigma'}^{b}c_{i\sigma}^{b},
\end{aligned}
\end{equation}
with Hubbard interaction, $U$, and Hund's coupling, $J_H$, where $V=U-2J_H$, and where $a$ and $b$ represent the $t_{2g}$ orbitals ($yz,xz,xy$). 
This can be expressed in terms of pairing order parameters, including the OSST parameters, which appear as,
\begin{align}
\frac{H_\text{eff}}{2N} = (V-J_H)\sum_{\nu} \hat{\textbf{D}}_\nu^\dagger(\textbf{q}) \cdot \hat{\textbf{D}}_\nu(\textbf{q}),
\end{align}
where $\hat{D}_\nu^{l\dagger}$ is given by,
\begin{equation}
\hat{D}_\nu^{l\dagger}(\textbf{q})=\frac{1}{4N}\sum_{\bf k}c_{{\bf k}\sigma}^{a\dagger}[i\hat{\sigma}^y \hat{\sigma}^l]_{\sigma \sigma'}[\hat{\lambda}_\nu]_{ab} c_{-{\bf k}+{\bf q}\sigma'}^{b\dagger},
\end{equation}
with $l=x,y,z$.
 ${\hat \lambda}_\nu$ are $3 \times 3$ anti-symmetric matrices in the orbital basis under the exchange
of the $t_{2g}$ orbitals for three different inter-orbital matrices denoted with $\nu = X$ (between $xz$ and $xy$ orbitals), $Y $($yz$ and $xy$), $Z$ ($xz$ and $yz$).
Their expressions are given in SM.  
The full form of the interaction written in terms of pairing order parameters, including induced intra-orbital spin-singlets $\phi_a$, and inter-orbital-triplet spin-singlets, both of which appear with repulsive interactions, are also given in the SM.
The OSST channel has an attractive interaction for $3J_H>U$, and while this is larger than most values of Hund's coupling in 4$d$ transition metals, where $J_H$ is about 20-30\% of $U$ \cite{Georges2013ARCMP}, recent studies going beyond mean-field theory support OSST pairing originating from Hund's coupling without the strict condition of $3J_H>U$ \cite{Hoshino2015PRL,Hoshino2016PRB,Gingras2019PRL,Kugler2019arXiv}. 
The direction of the d-vector is determined by the SOC \cite{Puetter2012EPL, Suh2019}, with order parameters belonging to the $A_{1g}$ representation for atomic SOC \cite{Ramires2019PRB,Huang2019PRB,Cheung2019PRB}.
The importance of the SOC in Sr$_2$RuO$_4$ was addressed earlier \cite{Pavarini2006PRB,Haverkort2008PRL,Rozbicki2011JPCM,Puetter2012EPL,Veenstra2014PRL}, and recently re-emphasized \cite{Tamai2019PRX}.
 
\begin{figure}
\includegraphics[width=0.9\columnwidth]{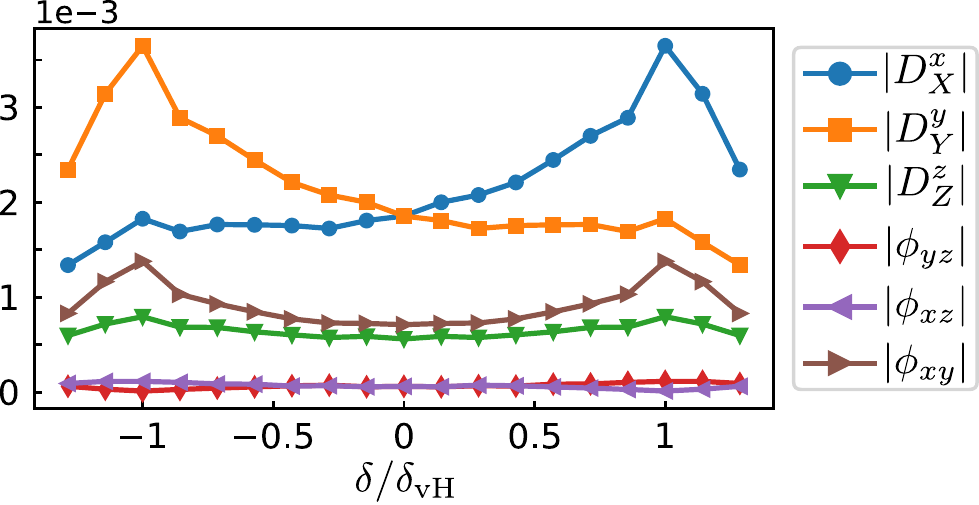}
\caption{Magnitude of the finite components of $\textbf{D}_\nu$ and $\phi_a$ as a function of strain $\delta$ showing roughly quadratic behaviour of the gap size in $\delta$ with a maximum at the vHS as expected.
Note that $D_X^x/D_Y^y$ and $\phi_{xz}/\phi_{yz}$ show the expected asymmetry with respect to $\pm \delta$.
}\label{strain}
\end{figure}

{\it Pairing gap under strain} --  
Since the OSST pairing corresponds to pairing between orbitals with different energies at $\textbf{k}$ and $-\textbf{k}$,
we consider the possibility
of finite momentum pairing, i.e., FFLO state.
Using a self-consistent mean field theory, we find the zero-momentum ${\bf q}=0$ state is always the lowest state despite the pairing between different orbitals. 
However, the pairing amplitude appears to be extremely small as shown in Fig.~\ref{strain} with the magnitudes of the $\textbf{D}_\nu$ and induced intra-band spin-singlets, $\phi_a$. 
They are thousands of times smaller
 than the t$_{2g}$ bandwidth, even though the attractive interaction is reasonably large. We set 
 $3J_H-U = 0.5$ for the current results, and the mean field theory in general overestimates the gap size. 
The inter-orbital pairing would appear to require a finite q value to produce a gap on the FS without orbital hybridization or SOC.  
However, when the atomic SOC is finite 
the OSST pairing projected onto the band basis transforms into \textit{intra-band} pseudospin-singlet pairing on the FS,
denoted by $\tilde{D}_i$ where $i = \alpha, \beta, \gamma$ in the quasiparticle dispersion shown in Fig.~S1.
 The quasiparticle dispersion represents strongly anisotropic gaps, which are very small in size, both at and below
 the FS.  This suggests that when the bandwidth is renormalized by electronic correlations, and becomes narrower, the OSST is further favoured. A  recent dynamical mean field theory reported a strong mass renormalization of the bands \cite{Kim2018PRL},
 which would also enhance the OSST pairing. 

To study the uniaxial strain effects, we change the ratio of the hopping integrals along the $a$- and $b$-axes such that
$ t_{jx}=(1- \delta)t_j$ and $t_{jy}=(1 + \delta)t_j$ for $j=1,2,3$. Uniaxial strain along the $a$-axis corresponds to $\delta < 0$. 
The change of different order parameters as a function of $\delta$ is shown in Fig.~\ref{strain}. The pairing gap is roughly quadratic in $\delta$ as expected from the even parity pairing.
While $D^x_X$ and $D^y_Y$ exhibit opposite behaviour under strain, these $A_{1g}$ solutions do not exhibit a split transition under strain \cite{Ramires2019PRB}.
When the $\gamma$ band touches the vHS around $\delta =\pm 0.07$, the pairing amplitude is peaked.
Since mean field theory causes the gap to be proportional to the transition temperature, $T_c$ is also peaked as reported in Refs.~\onlinecite{Hicks2014Science,Steppke2017Science}.
 The overall gap size is minuscule in comparison to the energy scale of the kinetic and potential terms as discussed above.

{\it Rotation of the d-vector under uniaxial strain} -- For spin-triplet pairing, the d-vector represents the direction along which the spin projection of the condensed pair has eigenvalue zero \cite{Leggett1975RMP}.
When SOC is finite, the mean field solutions find the pinning of the d-vector depending on the inter-orbital composition via SOC.
For the pairing between $xz$ and $xy$ orbitals, the d-vector points along the $x$-direction (represented by $D_X^x$), $yz$ and $xy$ along the $y$-direction ($D_Y^y$), and $xz$ and $yz$ along the $z$-axis ($D_Z^z$).
The $x$-,$y$-, and $z$-axes are the same as the crystallographic axes of $a$, $b$, and $c$, as Sr$_2$RuO$_4$ is a tetragonal lattice.
The d-vector changes in momentum space as shown in Fig.~\ref{fs}(a), as the orbital composition changes along the FS.
The red arrows represent the d-vector directions. The shorter the length of arrow, the bigger the $c$-axis component of the d-vector. 
There is a finite d-vector at every momentum point, and on average it is finite in all directions leading to a reduction of the spin polarization in all directions. 

In the absence of strain, due to the tetragonal symmetry, there is a $\frac{\pi}{2}$ rotational symmetry between ${\hat D}^x$ and ${\hat D}^y$.
This leads to the same reduction of the spin polarization along the $a$- and $b$-axes (and any other directions related to the symmetry of the tetragonal lattice). 
However, when the uniaxial strain is applied, the orbital composition changes mainly around X and Y regions of the Brillouin Zone (BZ) as shown
by the underlying FS in Fig.~\ref{fs}(b). Most importantly, the $yz$ orbital contribution to all bands increases, causing the d-vector at every momentum to
rotate towards the $b$-axis, with the most change occurring around the diagonal direction of the BZ.
This will then affect the magnitude of the spin polarization in the superconducting state, and generates a directional dependence, which
we show below. 

\begin{figure}
\includegraphics[width=0.9\columnwidth]{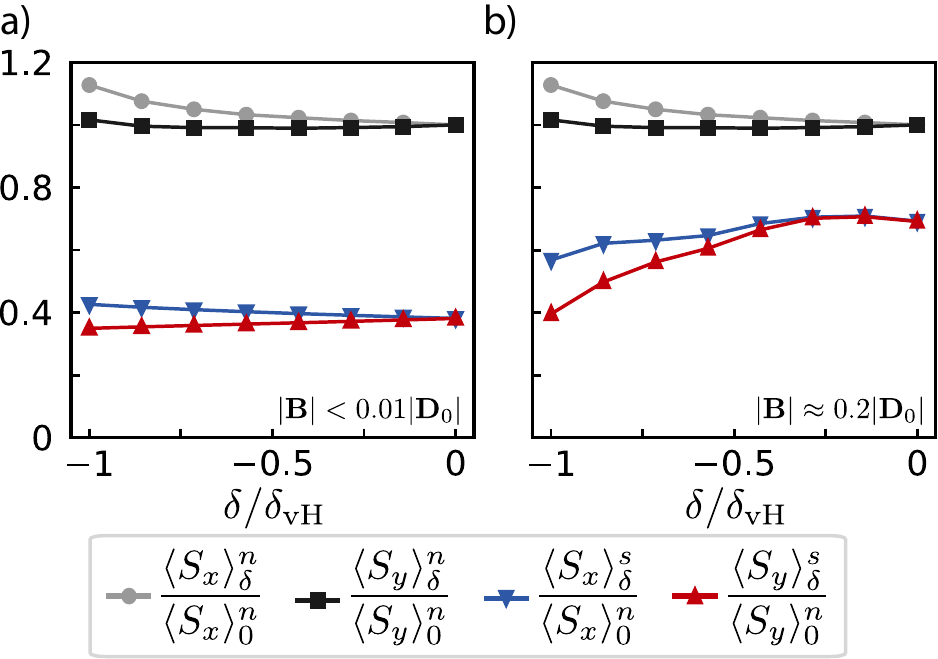}
\caption{
The spin magnetization $\textbf{S}$ in the normal ($n$) and superconducting ($s$) states, normalized to the zero-strain normal state value, for (a) a small field where $\textbf{S}$ is linear in $B$ with $B<1\%$ of $|\textbf{D}_0|$ where $|\textbf{D}_0|$ is $|\textbf{D}|$ at $\delta=0$,
and (b) a field comparable to the gap minimum, where $\textbf{S}$ is no longer linear, $B \approx 0.2 \times |\textbf{D}_0|$.
}\label{mag}
\end{figure}

{\it Spin polarization under strain} -- The magnetic susceptibility $\chi_{jj}$ measured by the NMR Knight shift is given by
$\partial M_j/\partial B_j$ where ${\bf M}$ is the magnetization, ${\bf B}$ is an external magnetic field, and $j = x, y, z$.
Using a Zeeman coupling $H_{\text{Zeeman}} =  \sum_i ( {\bf L}_i +  g {\bf S}_i) \cdot {\bf B}$, we compute the contribution from the spin polarization at a site $i$, in the $j$ direction,
$\langle S_i \rangle_j$, assuming the orbital contribution, which has been suggested to be small \cite{Ishida2019}, can be separated.
We also compute the contribution from the orbital magnetization $\langle L_i \rangle$, and there is a slight drop in the superconducting state as shown in Fig.~S2 in the SM.
  The results are shown in Fig.~\ref{mag}, which shows the spin magnetization along the $x$- and $y$-directions as the strain changes.
  Here we plot the ratio between the strained values, and the normal state unstrained cases.
  
A conventional spin-triplet will feature a Knight shift which appears the same as a singlet for the field parallel to the d-vector and shows no change from the normal state for the field perpendicular to the d-vector. 
On the other hand, OSST pairing leads to 
intra-band pseudospin-singlet pairing occuring near the FS, and inter-band spin-triplet away from the FS.
Thus, the low field response behaviour is due primarily to the intra-band pairing \cite{Yu2018PRB}, which causes a large drop in the approximately isotropic Knight shift as shown in Fig.~\ref{mag}a.
However, by increasing the field such that it is a significant fraction of the gap size $(\textbf{B}\sim 0.2 |\textbf{D}_0|)$, the inter-band pairing with d-vector rotation is observable, and such rotation results in an anisotropic Knight shift under strain as shown in Fig.\ref{mag}b.
Thus for OSST, the Knight shift is more affected by intra-band pseudospin-singlet pairing at low fields, and inter-band pairing at higher fields. 
As expected from the d-vector rotation under the $a$-axis strain,
we find a greater drop in the magnetization from the normal to superconducting state 
in the $y$-direction compared with the $x$-direction, with a difference of about $20\%$ for the larger field value. 
The magnetization in the $x$-direction also drops under strain due to the strain bringing the sample deeper into the superconducting state.  
The value of the drop from the normal to superconducting state depends on the value of the SOC, and by decreasing the SOC, the Knight shift drop and the anisotropy under strain enhance further.

\begin{figure}
 \includegraphics[width=1.0\columnwidth]{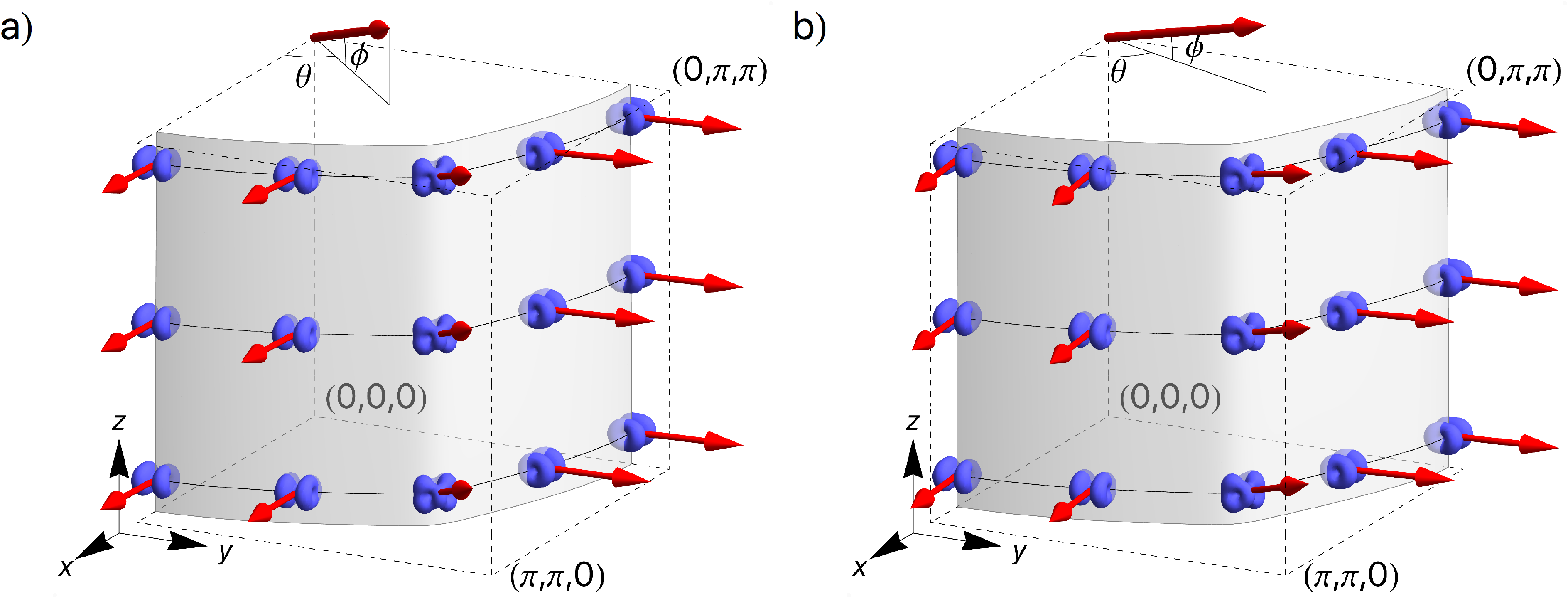}
 \caption{The d-vectors on the 3D $\gamma$-band FS are shown at various momentum points for (a) no strain and (b) strain along the
 $a$-axis.  The average d-vector indicated by the red arrow at the top corner shows the rotation of the d-vector denoted by $(\theta, \phi)$
  towards $b$-axis  and slightly $c$-axis under the $a$-axis uniaxial strain. Similar to the 2D case, most of the rotation of d-vector occurs
  near the diagonal direction of the BZ. 
 }\label{3d} 
 \end{figure}

 {\it  Extending to three-dimensional bands} -- 
 Sr$_2$RuO$_4$ has a layered structure, and one expects to see more $k_z$ dispersion of the bands originating from $xz$ and $yz$ orbitals due to their shape, and less dispersion from the $xy$ orbital. 
 The momentum dependent $t_{2g}$-orbital projection of the wavefunction for the $\alpha$, $\beta$ and $\gamma$
 bands on the three-dimensional FS was reported \cite{Veenstra2014PRL}, which is consistent with the three dimensional (3D) tight binding model
 constructed in Ref.~\onlinecite{Roising2019prr}.
 The $\beta$ and $\gamma$ bands still have significant overlap of $xy$ and one dimensional (1D) orbitals,
 even though detailed composition depends on $k_z$ as shown in Ref.~\onlinecite{Veenstra2014PRL}, while the $\alpha$ band is mainly made of 1D orbitals. Thus the above analysis done in the two-dimensional  (2D) system can be generalized
 to a layered three-dimensional system. The qualitative uniaxial strain effect, i.e.,  the relative directional dependence of 
 the spin polarization under a uniaxial strain, is independent of the details of $c$-axis hopping parameters, even though the quantitative drop may depend
 on the strength of the hopping parameters. 
 Using the tight binding parameters in Ref.~\onlinecite{Roising2019prr}, we found the d-vector directions are similar to the 2D case.
 The angle $\phi$ represents the tilting from the $ab$-plane, which is about $17-19^\circ$ depending on $k_z$. 
 A clear rotation of the averaged d-vector is shown as a red arrow in a top corner in Fig.~\ref{3d}, and the main conclusion of the d-vector rotation 
 can be generalized to the 3D model including the layer coupling. 

{\it Discussion and summary} --
In multi-orbital systems, orbital degrees of freedom extend the types of superconducting pairings. Even-parity spin-triplet pairings are allowed when the pairing
occurs between different orbitals with the antisymmetric fermionic wavefunction condition, i.e., orbital singlets.
In the band basis, this maps to inter-band, and intra-band pairings when SOC is finite \cite{Puetter2012EPL}. 
SOC in the OSST pairing determines the intra-band gap on the FS. 
The idea of OSST pairing is not limited to the atomic (s-wave) SOC leading to an s-wave gap with the $A_{1g}$ representation.  It can be generalized to the momentum dependent SOC terms resulting in higher order OSST terms such as d- and g-wave, which can explain the nodal structure of the gap, as well as recent experiments suggesting a multi-component order parameter based on the elastic moduli \cite{Benhabib2020,Ghosh2020}.  These experiments have contributed to a recent proposal of a combined d- and g-wave gap \cite{Kivelson2020npj}.  
Within the OSST pairing scenario, the multi-component order parameter can be found using the momentum dependent SOC including d- and g-wave SOC terms. 
We find a finite d- and g-wave gap structure with the d- and g-wave SOC terms as expected. However, a self-consistent solution to determine the energetics of all possible pairings is the subject of future work. 


Odd-parity, intra-orbital (and inter-orbital triplet) spin-triplet pairing is also possible when the ferromagnetic interaction is extended to further neigbour site \cite{Ramires2019PRB},
even though the impact of the increased density of states via the vHS is drastically reduced due to the $\sin (nk_{x/y})$ form factor, where $n$ is an integer representing the nearest and further nearest neighbour distance.
An interaction of this form has been shown to give a helical order parameter \cite{Rice1995JPCM,Ng2000EPL,Mackenzie2003RMP}. 
We consider this possibility and show that there is still potentially an asymmetry of the in-plane magnetic response for an in-plane d-vector in the SM.
While recent neutron scattering experiments and DMFT calculations do not support ferromagnetic spin fluctuations giving rise to spin-triplet pairing \cite{Steffens2019PRL,Strand2019PRB}, this triplet leads to an anisotropy, but cannot reproduce the $>50\%$ drop in the NMR response, with a $\sim40\%$ or smaller drop in the NMR response for all field strengths, unlike the OSST pairing.
For this odd-parity pairing, the drop in the Knight shift with no SOC is $50\%$, and the inclusion of SOC decreases the magnitude of the drop. 
Therefore, if an anisotropy in the Knight shift is observed, the odd-parity spin-triplet solutions must also be considered as possible explanations, however, the exact value of the drop in the Knight shift will provide important evidence for identifying the pairing states.
For unstrained samples, the impact of pulse energy is stronger than in the case of strained samples as stated in Ref.~\onlinecite{Pustogow2019Nature},
and further experimental analysis is required to determine if the reduction is more than 50\% in the unstrained case.

Another consequence of SOC is a complex order parameter. 
Generally, the order parameter with the SOC induced spin-singlet components can be written with a phase factor, $\hat{\textbf{D}} + e^{i \theta} \hat\phi$ (where $\hat{\textbf{D}}$ is defined to be imaginary such that it is even under TR), where the relative phase between the two is determined by the atomic SOC \cite{Puetter2012EPL}, and $\theta=0$ for uniform SOC.
 Despite not breaking time-reversal symmetry, as the time-reversal operator maps the order parameter to
 itself, 
 the order parameter near impurities may change its relative phase from $\theta=0$ leading to non-trivial effects.
Thus, the multi-component order parameter may be important to 
understand the $\mu$SR \cite{Luke1998Nature} and Josephson junction \cite{Kashiwaya2019prb} results.
This is an open topic for future study.

In summary, we showed 
OSST pairing with SOC leads to a significant reduction of the Knight shift and an anisotropic Knight shift response under uniaxial strain, which can ultimately be used to differentiate spin-triplet from spin-singlet pairing in Sr$_2$RuO$_4$.
When the strain is applied along the $a$-axis, inter-orbital pairing involving d$_{yz}$ and d$_{xy}$ is further enhanced 
 leading to a rotation of the d-vector towards the $b$-axis. As a consequence of the d-vector rotation, the Knight shift
becomes anisotropic relative to the strain axis. 
It has more drop in the magnetization when the magnetic field is perpendicular to the strain and less when the field is parallel to the strain. 
Such anisotropy is not expected in the spin-singlet, thus we propose the Knight shift measurement
with the field along the $a$-axis, which can be compared with the data presented in Ref.~\onlinecite{Pustogow2019Nature}. This will ultimately determine
a long-standing debate of a possible spin-triplet pairing in Sr$_2$RuO$_4$. This idea can also be extended to other multi-orbital systems with significant
Hund's coupling and SOC.

\begin{acknowledgments} 
This work was supported by the Natural Sciences and Engineering Research Council of Canada Discovery Grant 06089-2016, and the Center for Quantum Materials at the University of Toronto.
\end{acknowledgments} 

\bibliography{Sr2RuO4-NMR}

{
\newpage
\onecolumngrid
\appendix
\setcounter{secnumdepth}{2}
\section{Hamiltonian}

\renewcommand{\thefigure}{S\arabic{figure}}
\renewcommand{\theequation}{S\arabic{equation}}
\renewcommand{\thetable}{S\Roman{table}}

The two-dimensional tight binding and SOC Hamiltonians referenced in the main text are given by
\begin{equation}\label{kin}
H_{\text{kin}}= \sum_{k,\sigma,a} \epsilon_k^a c_{k\sigma}^{a\dagger} c_{k\sigma}^a + \sum_{k,\sigma} \epsilon_k^{1d}  c_{k\sigma}^{yz\dagger} c_{k\sigma}^{xz} + \text{h.c.},
\end{equation}
\begin{equation}\label{soc}
H_{\text{SOC}}=i\lambda \sum_i \sum_{abl} \epsilon_{abl} c_{i\sigma}^{a\dagger}c_{i\sigma'}^b \hat{\sigma}_{\sigma\sigma'}^l,
\end{equation}
where the dispersions for the kinetic terms in Eq.~\ref{kin} are given by $\epsilon_k^{yx/xz} = -2t_1 \cos k_{y/x}-2t_2\cos k_{x/y} -\mu_1$, $\epsilon_k^{xy} = -2t_3(\cos k_x+\cos k_y) - 4t_4 \cos k_x \cos k_y - \mu_2 $, with the parameters listed in Table \ref{tb}.
\begin{table}[h]
\centering
\begin{tabularx}{0.5\columnwidth}{>{\centering}X >{\centering}X >{\centering}X >{\centering}X >{\centering}X >{\centering}X >{\centering}X >{\centering}X}
    $t_1$ & $t_2$ & $t_3$ & $t_4$ & $t_5$ & $\lambda$ & $\mu_1$ & $\mu_2$ \tabularnewline \hline
    $0.45$ & $0.05$ & $0.5$ & $0.2$ & $0.025$ & $0.085$ & $0.531$ & $0.631$
\end{tabularx}
\caption{Tight binding parameters used in Eqs.~\ref{kin} and \ref{soc} in the main text.}\label{tb}
\end{table}

The full interacting Hamiltonian, Eq.~(1), rewritten in terms of the pairing order parameters, is given by,
\begin{align}
\frac{H_\text{int}}{2N}  &= 2U \sum_{a} \hat{\phi}_a^\dagger(\textbf{q})\hat{\phi}_a (\textbf{q}) + (V-J_H)\sum_{\nu} \hat{\textbf{D}}_\nu^\dagger(\textbf{q}) \cdot \hat{\textbf{D}}_\nu(\textbf{q}) \nonumber \\ &+ 2J_H\sum_{a\ne b}\hat{\phi}_a^\dagger(\textbf{q})\hat{\phi}_b (\textbf{q}) + (V+J_H)\sum_{\nu}\hat{\phi}_\nu^\dagger(\textbf{q})\hat{\phi}_\nu (\textbf{q}),
\end{align}
with the intra-orbital, spin-singlet parameter,
\begin{equation}
\hat{\phi}_{a}^\dagger (\textbf{q}) = \frac{1}{4N}\sum_{\bf k} c_{{\bf k}\sigma}^{a\dagger}[i\hat{\sigma}^y]_{\sigma \sigma'}c_{-{\bf k}+{\bf q}\sigma'}^{a\dagger},
\end{equation}
and the inter-orbital spin-singlet order parameter, 
\begin{equation}\label{phi}
\hat{\phi}_{\nu}^\dagger (\textbf{q}) = \frac{1}{4N}\sum_{\bf k} c_{{\bf k}\sigma}^{a\dagger}[i\hat{\sigma}^y]_{\sigma \sigma'}[\hat{\epsilon}_\nu]_{ab}c_{-{\bf k}+{\bf q}\sigma'}^{b\dagger},
\end{equation}
as well as the inter-orbital-singlet, spin-triplet, parameter given in Eq.~(3) of the main text.
To express the inter-orbital spin-triplet and spin-singlet order parameters with orbital degrees of freedom, we use the $3 \times 3$ Gell-Mann matrices in the orbital
basis. ${\hat \lambda}_\nu$ and ${\hat \epsilon}_\nu$ in Eq.~(3) of the main text and Eq.~\ref{phi} are defined by
\begin{equation}
\hat\epsilon_{X} =  
\begin{pmatrix}
	0 & 0 & 0\\
	0 & 0 & 1\\
	0 & 1 & 0
\end{pmatrix}
,\hat\epsilon_{Y} =  
\begin{pmatrix}
	0 & 0 & 1\\
	0 & 0 & 0\\
	1 & 0 & 0
\end{pmatrix}
,\hat\epsilon_{Z} =  
\begin{pmatrix}
	0 & 1 & 0\\
	1 & 0 & 0\\
	0 & 0 & 0
\end{pmatrix}
\end{equation}
\begin{equation}
\hat\lambda_X =  
\begin{pmatrix}
	0 & 0 & 0\\
	0 & 0 & i\\
	0 & -i & 0
\end{pmatrix}
,\hat\lambda_Y =  
\begin{pmatrix}
	0 & 0 & -i\\
	0 & 0 & 0\\
	i & 0 & 0
\end{pmatrix}
,\hat\lambda_Z =  
\begin{pmatrix}
	0 & i & 0\\
	-i & 0 & 0\\
	0 & 0 & 0
\end{pmatrix}
\end{equation}

\section{Quasiparticle Dispersion}

Given that the inter-orbital pairing occurs among $t_{2g}$ orbitals, which form different bands with different Fermi momenta, the pairing gap is finite not only near the FS, but also below the Fermi level.
For example, by finding the energy eigenvalues of the mean field Hamiltonian using the tight binding parameters listed above, as well as self consistent solutions for the various gap parameters, the quasiparticle dispersion is plotted in Fig.~\ref{bs}.  When a particle $\beta$-band intersects a hole $\gamma$-band at a finite energy above or below the FS,
a finite gap of $|D_\nu|$ is clearly present, as shown by the red circles in Fig.~\ref{bs}(b).  When a particle band crosses its own hole band at the FS, a gap will form due to intra-band pseudospin-singlet pairing $\tilde{D}_i$ with $i=\alpha,\beta,\gamma$, such as those labelled by the green circles in Fig.~\ref{bs}(b).  
\begin{figure}
\includegraphics[width=1.0\columnwidth]{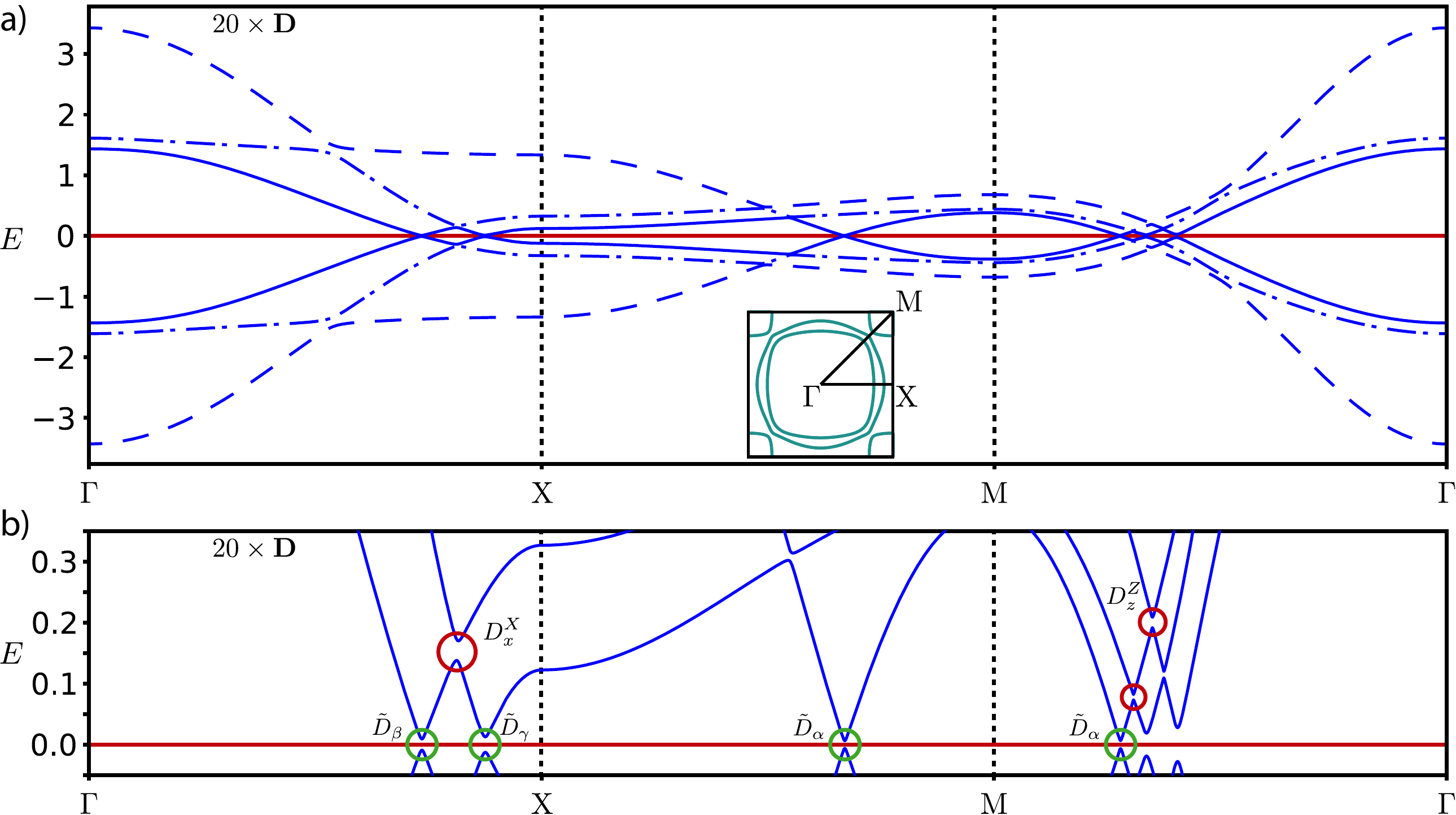}
\caption{(a) Bogoliubov quasiparticle dispersion, featuring bands of mixed particle and hole character, obtained by finding the energy eigenvalues of the Hamiltonian listed in the main text solved in the mean field approximation with $\phi$ and $D$ values obtained by self-consistent mean field theory for $3J-U = 0.5$ along the path shown on the inset Fermi surface.  
The gap parameters are increased by a factor of $20\times$ to make the
gaps more visible.  (b) Magnification of (a) to show the pairing occurring away from the
Fermi energy due to the inter-orbital nature of the triplet parameters, labelled with red circles.  Two of the red circles are labelled with the parameter that is primarily responsible for the pairing at that location, while a third is left unlabelled, where there will be mixed contribution due to the mixed orbital character of the bands around this location.  
Green circles show three locations of the intra-band pseudospin-singlet pairing (represented by $\tilde{D}_i$ with $i=\alpha,\beta,\gamma$) at the Fermi energy due to the OSST being projected onto the bands via the atomic SOC.
A tiny induced spin-singlet, $\phi_a$, also contributes to the intra-band pairing $\tilde{D}$.
The dispersion is shown in the unstrained case, and therefore the $\frac{\pi}{2}$ rotational symmetry means that the pairings due to $D^Y_y$ along $\Gamma$ to $\text{Y}$ is identical to $D^X_x$ along $\Gamma$ to $\text{Y}$.
}\label{bs}
\end{figure}
The gap on the FS is finite at every momentum, even though it is strongly anisotropic because of different orbital
composition in each band. 
This is due to the projection of the OSST pairing onto the band-basis \cite{Puetter2012EPL}. 
The gap at the FS ranges from approximately $0.1 |\textbf{D}_0|$ to $0.5|\textbf{D}_0|$ and is smallest in the $\alpha$ near the Brillouin zone boundary.  This depends on the size of the SOC, and decreasing the value of the SOC will decrease the size of the gap.  Additionally, a momentum dependent SOC can lead to gap nodes \cite{Suh2019}.  
While only the s-wave gap is considered in the present study, there may be significant contributions from higher-angular momentum pairing such as d-wave, originating from further neighbour interactions, which could also contribute to the formation of a node in the gap.  One may wonder if a finite momentum pairing, i.e., FFLO state occurs. We found 
that zero-momentum pairing between different bands has lower energy than an FFLO state for all parameters that we have considered. 

\section{Magnetic Response}

To calculate the magnetic response in the superconducting and normal states, a magnetic field is introduced to the Hamiltonian of the form,
\begin{equation}
H_B = -\mu_B \vec{H} \cdot \sum_{i} \vec{L}_i + 2\vec{S}_i,
\end{equation}
and the value of the spin magnetization in different directions, $\langle S^\alpha \rangle$, is calculated with and without finite superconducting order parameters to compare the superconducting and normal states, where the field direction is parallel to the the computed component (i.e., only $H^\alpha$ is finite).
The low field calculations (Fig.~3a in the main text) are performed where the response is linear, while Fig.~3b is for field strengths a significant fraction of the gap size, where the response is no longer linear.
The orbital contribution to the total magnetization has been suggested to be small \cite{Ishida2019}. 
The calculations of the orbital contribution to the total magnetization as a function of strain are presented here in Fig.~\ref{lmag}.
This shows only a small change from the normal to superconducting state.  
Note that for a conventional singlet superconductor there is no change in the orbital magnetization from the normal to superconducting state.

\begin{figure}
\includegraphics[width=0.5\columnwidth]{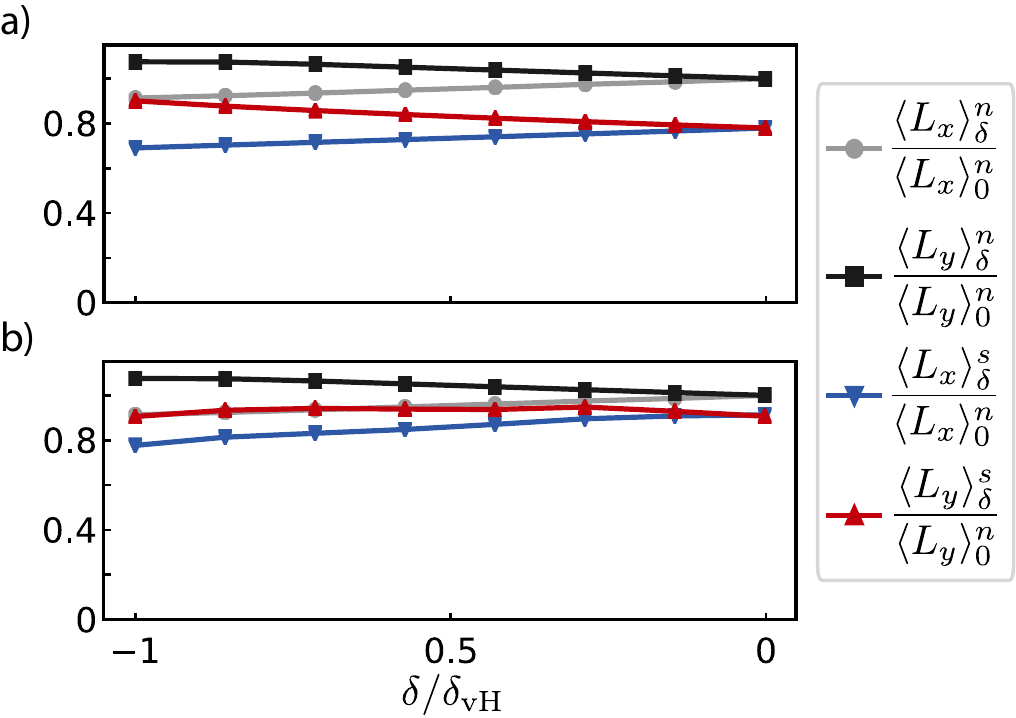}
\caption{Orbital magnetization $\langle \textbf{L} \rangle$ in the normal (n) and superconducting (s) states, normalized to the zero-strain, normal state value for (a) a small field where $\textbf{L}$ is linear in $\textbf{B}$ with $B<1\%$ of $|\textbf{D}_0|$ 
and (b) a field comparable to the gap minimum, where $\textbf{L}$ is no longer linear, $B \approx 0.2 |\textbf{D}_0|$.
}\label{lmag}
\end{figure}

\section{Odd-Parity Pairing}

To show that d-vector rotation can occur for other types of triplet pairing, an odd-parity order parameter is considered by including a nearest-neighbor intra-orbital ferromagnetic interaction,
\begin{equation}
H_{\text{int}} = -J\sum_{\langle i,j\rangle, a} \vec{S}_i^a \cdot \vec{S}_j^a,
\end{equation}
which gives rise to odd-parity spin-triplet terms with an attractive interaction, as well as repulsive even-parity spin-singlet terms. 
Considering only the attractive triplet terms,
\begin{equation}
\frac{H_\text{int}}{N} = \frac{-J}{2}\sum_a \textbf{d}_x^{a\dagger}\cdot \textbf{d}_x^{a}+\textbf{d}_y^{a\dagger}\cdot \textbf{d}_y^{a},
\end{equation}
where,
\begin{equation}
\textbf{d}^a_{x/y} = \frac{1}{2N}\sum_k c_{-k,\sigma}^a [\sigma_y i \vec{\sigma}]_{\sigma,\sigma'} c_{k,\sigma'}^a \sin k_{x/y}.
\end{equation}
While inter-orbital terms can also be considered, which can give rise to even-parity spin-triplet with an attractive interaction, as well as odd-parity spin-singlet terms, we limit this calculation to only the intra-orbital terms, all of which are chosen to be equal.
For an interaction twice the size of the interaction considered for the $A_{1g}$ order parameter calculations, 
$J=2(3J_H-U)=1$, with the same two-dimensional tight-binding parameters, helical solutions can be found of the form 
$\textbf{d} = \Delta_x \sin k_y \hat{x} - \Delta_y \sin k_x \hat{y}$ or 
$\textbf{d} = \Delta_x \sin k_x \hat{x} + \Delta_y \sin k_y \hat{y}$ for each of the orbitals, both belonging to the $E$ representation \cite{Ramires2019PRB}, where $|\textbf{d}|\approx |\textbf{D}|$.
Under uniaxial strain, the component of the d-vector with the $\sin k_x$ dependence becomes larger, which leads to an anisotropy in the magnetic response, as shown in Fig.~\ref{pmag}.
While the two helical solutions found are degenerate, and therefore a combinations of the two states could cause the d-vector to rotate depending on the field, the degeneracy can be broken by an inter-orbital interaction \cite{Ng2000EPL}, which could lead to the anisotropic response shown here.
In both cases, the drop is $\sim40\%$ or smaller, independent of the field strength, which cannot reproduce the experimentally observed $>50\%$ drop in the NMR response. 
\begin{figure}
\includegraphics[width=0.5\columnwidth]{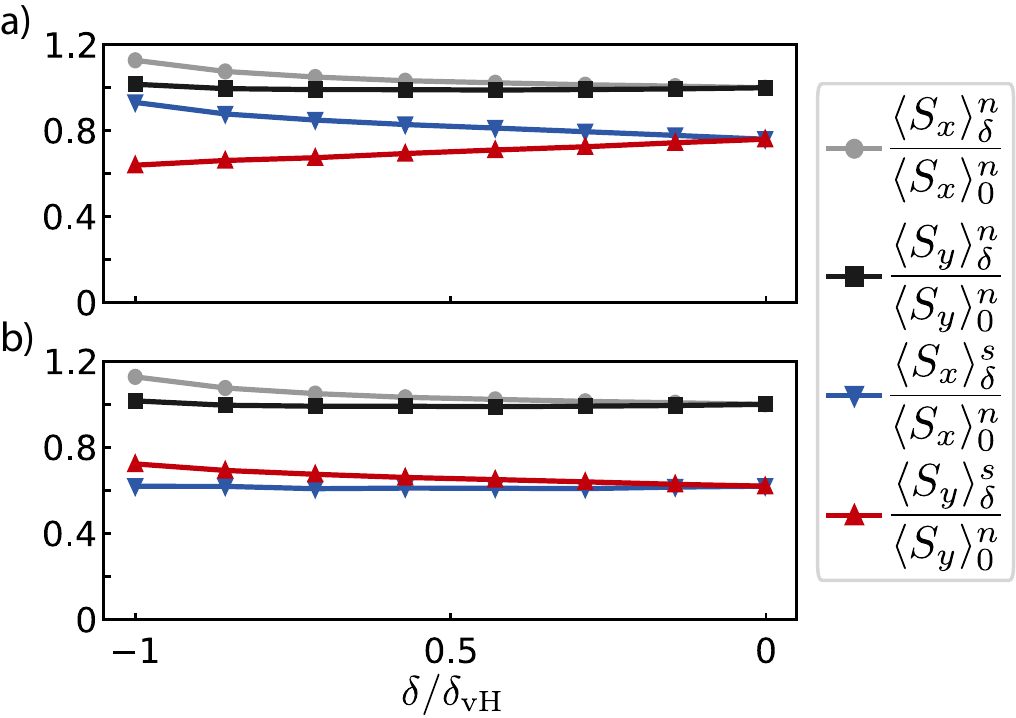}
\caption{Spin component of the magnetization under strain for the (a) $\Delta_x \sin k_y \hat{x} - \Delta_y \sin k_x \hat{y}$ and (b) $\Delta_x \sin k_x \hat{x} + \Delta_y \sin k_y \hat{y}$ p-wave solutions, for a field much less than the gap size $|\textbf{B}|<1\%$ of $|\textbf{D}_0|$.
The higher field magnetization is similar to the low field magnetization shown here.
In both cases, there is a decrease in the Knight shift component for the d-vector direction with the $\sin k_x$ dependence, since this component increases under uniaxial strain along the a-axis.
}\label{pmag}
\end{figure}


}

\end{document}